\documentstyle[12pt]{article}

\catcode`\@=11
\begin{document}
\centerline{\large\bf The Evolution Operator of The Three Atoms
Tavis-Cummings Model  }
 \vspace{0.8cm}
 \centerline{\sf Jin-fang Cai} \baselineskip=13pt \vspace{0.5cm}
\centerline{Department of Applied Physics, Beijing Institute of
Technology, }
 \baselineskip=12pt
 \centerline{ Beijing, 100081, P.R. China }
\vspace{0.9cm}
\begin{abstract} {The explicit form of evolution operator of the three atoms 
Tavis-Cummings Model is given. }
\end{abstract}
\vspace{1.2cm}

The Tavis-Cummings Models have been widely studied in the field of
quantum optic and quantum computer. In this brief letter, The
analytical form of the evolution operators of T-C model with three
atoms is given. Compared with the result in the paper\cite{fujii},
our result is more convenient for further application.

Under the rotating-wave approximation and resonant condition, the
Hamiltonian of the Tavis-Cummings model with three atoms in the
interaction picture is
\begin{equation}\hat{H}_{I}=\hbar\gamma\sum_{i=1}^3(\hat{a}\hat{\sigma}_i^++\hat{a}^{\dag}\hat{\sigma}_i^-)
\end{equation}
We denote the ground and exited states for the atom by $|g\rangle$
and $|e\rangle$. The evolution operator
$\hat{U}(t)=\exp(-i\hat{H}t/\hbar)$ in the atomic basis
$|eee\rangle,|eeg\rangle,|ege\rangle,|gee\rangle,|egg\rangle,|geg\rangle,|gge\rangle,|ggg\rangle,$
is found to be \begin{equation} \left(
\begin{array}{ccc}
  \hat{U}_{11} & \cdots & \hat{U}_{18} \\
  \vdots & \ddots & \vdots \\
  \hat{U}_{81} & \cdots & \hat{U}_{88} \\
\end{array}%
\right)\end{equation}
 where
 \begin{eqnarray}
\hat{U}_{11} &=&\frac{(7+2\hat{N}+\hat{\Omega})\cos
(\hat{\Theta_1}\gamma t)+(-7-2\hat{N}+\hat{\Omega})\cos
(\hat{\Theta_2}\gamma t )}{2\hat{\Omega}
 }\equiv u_{11}(\hat{N})\nonumber \\
 \hat{U}_{22} &=&\hat{a}^{\dag}\frac{\left( -1 - 2\hat{N} + \hat{\Omega}  \right) \,\cos (\hat{\Theta_1}\,\gamma t )
 + \left( 1 + 2\hat{N} + \hat{\Omega}  \right) \,\cos (\hat{\Theta_2}\gamma t ) +
    4\,\hat{\Omega} \,\cos ({\sqrt{2 +\hat{N}}}\,\gamma t )}{6\,\hat{\Omega}(\hat{N}+1) }\hat{a}
 \nonumber\\   & \equiv&\hat{a}^{\dag}\frac{u_{22}(\hat{N})}{\hat{N}+1}\hat{a}\nonumber \\
 \hat{U}_{88}&=&\hat{a}^{\dag 3}\frac{\left( 1 + 2\hat{N} + \hat{\Omega}  \right) \,\cos
(\hat{\Theta_1}\gamma t ) + \left( -1 - 2\hat{N} + \hat{\Omega}
\right) \,\cos (\hat{\Theta_2}\gamma t )}{2\,\hat{\Omega} (\hat{N}+1)(\hat{N}+2)(\hat{N}+3)}\hat{a}^3\nonumber \\
    & \equiv&\hat{a}^{\dag 3}\frac{u_{88}(\hat{N})}{(\hat{N}+1)(\hat{N}+2)(\hat{N}+3)}\hat{a}^3\nonumber \\
\hat{U}_{12}&=&\frac{ \hat{\Theta_1}\,\left( 7 + 2\hat{N} +
\hat{\Omega} \right) \,\sin (\hat{\Theta_1}\gamma t ) +
 \hat{\Theta_2}\,\left( -7 - 2\hat{N} + \hat{\Omega}  \right) \,\sin (\hat{\Theta_2}\gamma t )  }{6i{(1+\hat{N})}\,
 \hat{\Omega} }\hat{a}
 \equiv \frac{u_{12}(\hat{N})}{\sqrt{\hat{N}+1}}\hat{a}\nonumber \\
\hat{U}_{15}&=&\frac{ -\cos (\hat{\Theta_1}\gamma t ) + \cos
(\hat{\Theta_2}\gamma t )  }
  {\hat{\Omega}}\hat{a}^2
  \equiv\frac{u_{15}(\hat{N})}{\sqrt{(\hat{N}+1)(\hat{N}+2)}}\hat{a}^2\nonumber \\
 \hat{U}_{25}&=&\hat{a}^{\dag}\frac{-\hat{\Theta_1}\,(2+\hat{N})
\sin (\hat{\Theta_1}\gamma t ) +\hat{\Theta_2}\,(2+\hat{N}) \,\sin
(\hat{\Theta_2}\gamma t ) + {\sqrt{2 +\hat{N}}}\,\hat{\Omega}
\,\sin ({\sqrt{2 +\hat{N}}}\,\gamma t )}{3i(1 +\hat{N})(2 +\hat{N})\,\hat{\Omega} }\hat{a}^2\nonumber \\
&\equiv&
\hat{a}^{\dag}\frac{u_{25}(\hat{N})}{(\hat{N}+1)\sqrt{\hat{N}+2}}\hat{a}^2\nonumber\\
\hat{U}_{58}&=&\hat{a}^{\dag 2}\frac{ \hat{\Theta_1}\,\left( 1 +
2\hat{N} + \hat{\Omega} \right) \,\sin (\hat{\Theta_1}\gamma t ) +
      \hat{\Theta_2}\,\left( -1 - 2\hat{N} + \hat{\Omega}  \right) \,\sin (\hat{\Theta_2}\gamma t )}
      {6i{(1 +\hat{N})(2 +\hat{N})(3 +\hat{N})}\,\hat{\Omega} }\hat{a}^3\nonumber \\
&\equiv&
\hat{a}^{\dag 2}\frac{u_{58}(\hat{N})}{(\hat{N}+1)(\hat{N}+2)\sqrt{\hat{N}+3}}\hat{a}^3\nonumber\\
\hat{U}_{18}&=&\frac{i\hat{\Theta_1}\hat{\Theta_2}\left(
\hat{\Theta_2}\sin (\hat{\Theta_1} \gamma t ) - \hat{\Theta_1}
\sin (\hat{\Theta_2} \gamma t ) \right)
}{3(\hat{N}+1)(\hat{N}+3)\,\hat{\Omega} }\hat{a}^3\equiv
\frac{u_{18}(\hat{N})}{\sqrt{(\hat{N}+1)(\hat{N}+2)(\hat{N}+3)}}\hat{a}^3\nonumber
\end{eqnarray}
and
\begin{eqnarray}
&&\hat{U}_{33}=\hat{U}_{44}=\hat{U}_{22};\nonumber \\
&&\hat{U}_{55}=\hat{U}_{66}=\hat{U}_{77}=\hat{a}^{\dag
2}\frac{u_{22}(\hat{N})-
\frac{1}{\hat{\Omega}}(\cos (\hat{\Theta_1}\gamma t )-\cos (\hat{\Theta_2}\gamma t ))}{(\hat{N}+1)(\hat{N}+2)}\hat{a}^2\nonumber \\
&&\hat{U}_{23}=\hat{U}_{24}=\hat{U}_{32}=\hat{U}_{34}=\hat{U}_{42}=\hat{U}_{43}=
\hat{a}^{\dag}\frac{u_{22}(\hat{N})- \cos (\sqrt{\hat{N}+2}\gamma
t )}{\hat{N}+1}\hat{a} \nonumber
\\
&&\hat{U}_{56}=\hat{U}_{57}=\hat{U}_{65}=\hat{U}_{67}=\hat{U}_{75}=\hat{U}_{76}=\hat{a}^{\dag
2}\frac{u_{22}(\hat{N})- \frac{1}{\hat{\Omega}}(\cos
(\hat{\Theta_1}\gamma t )-\cos (\hat{\Theta_2}\gamma t
))-\cos(\sqrt{2+\hat{N}}\gamma t)}
{(\hat{N}+1)(\hat{N}+2)}\hat{a}^2 \nonumber\\
&&\hat{U}_{13}=\hat{U}_{14}=\hat{U}_{12};\hspace{1cm}
\hat{U}_{21}=\hat{U}_{31}=\hat{U}_{41}=\hat{a}^{\dagger}\frac{u_{12}(\hat{N})}{\sqrt{\hat{N}+1}}\nonumber
\\
&&\hat{U}_{15}=\hat{U}_{16}=\hat{U}_{17};\hspace{1cm}\hat{U}_{51}=\hat{U}_{61}=\hat{U}_{71}=\hat{a}^{\dagger
2}\frac{u_{15}(\hat{N})}{\sqrt{(\hat{N}+1)(\hat{N}+2)}}\nonumber\\
&&\hat{U}_{28}=\hat{U}_{38}= \hat{U}_{48}=
\hat{a}^{\dag}\frac{u_{15}(\hat{N})}
{(\hat{N}+1)\sqrt{(\hat{N}+1)(\hat{N}+2)}}\hat{a}^3;\nonumber\\
&&\hat{U}_{82}= \hat{U}_{83}= \hat{U}_{84}=\hat{a}^{\dag
3}\frac{u_{15}(\hat{N})}
{(\hat{N}+1)\sqrt{(\hat{N}+1)(\hat{N}+2)}}\hat{a}\nonumber\\
&&\hat{U}_{25}=\hat{U}_{26}=\hat{U}_{35}=\hat{U}_{37}=\hat{U}_{46}=\hat{U}_{47};\nonumber\\
&&\hat{U}_{52}=\hat{U}_{53}=\hat{U}_{62}=\hat{U}_{64}=\hat{U}_{73}=\hat{U}_{74}=
\hat{a}^{\dag
2}\frac{u_{25}(\hat{N})}{(\hat{N}+1)\sqrt{\hat{N}+2}}\hat{a}\nonumber\\
&&\hat{U}_{27}=\hat{U}_{36}=\hat{U}_{45}=
\hat{a}^{\dag}\frac{u_{25}(\hat{N})+\sin(\sqrt{2+\hat{N}}\gamma t)}{(1+\hat{N})\sqrt{2+\hat{N}}}\hat{a}^2; \nonumber\\
&&\hat{U}_{54}=\hat{U}_{63}=\hat{U}_{72}=\hat{a}^{\dag
2}\frac{u_{25}(\hat{N})+\sin(\sqrt{2+\hat{N}}\gamma t)}
{(1+\hat{N})\sqrt{2+\hat{N}}}\hat{a}\nonumber\\
&&\hat{U}_{68}=\hat{U}_{78}=\hat{U}_{58};\hspace{1cm}\hat{U}_{85}=\hat{U}_{86}=\hat{U}_{87}=\hat{a}^{\dag
3}\frac{u_{58}(\hat{N})}{(\hat{N}+1)(\hat{N}+2)\sqrt{\hat{N}+3}}\hat{a}^2\nonumber\\
&&\hat{U}_{81}=\hat{a}^{\dag
3}\frac{u_{18}(\hat{N})}{\sqrt{(\hat{N}+1)(\hat{N}+2)(\hat{N}+3)}}\nonumber
\end{eqnarray}
 where $\hat{\Omega},\hat{\Theta_1},\hat{\Theta_2}$ are
defined by
$$\hat{\Theta_1}^2=5(\hat{N}+2)-\hat{\Omega} \hspace{1cm} \hat{\Theta_2}^2=5(\hat{N}+2)+\hat{\Omega}$$
$$\hat{\Omega}^2=9+16(\hat{N}+2)^2$$
and $\hat{N}=\hat{a}^{\dag}\hat{a} $

\end{document}